\documentclass[a4paper,11pt]{article}
\usepackage{pos}
\renewcommand{\logo}{\relax}
\usepackage{graphicx}
\usepackage{wrapfig}
\newcommand{\SimProp}{\textit{SimProp}}
\newcommand{\Mpc}{\mathrm{Mpc}}
\newcommand{\EeV}{\mathrm{EeV}}
\newcommand{\etal}{et~al.\@}

\title{Simulating the propagation of cosmic rays heavier than iron in \SimProp}

\author*{Armando di Matteo}

\affiliation{Istituto Nazionale di Fisica Nucleare (INFN), Sezione di Torino,\\
  Via Pietro Giuria 1, 10125 Torino, Italy}

\emailAdd{armando.dimatteo@to.infn.it}

\abstract{Ultra-high-energy cosmic rays (UHECRs) have long been assumed to entirely consist of iron and/or lighter atomic nuclei, and this assumption has been hard-coded in a great deal of software for UHECR simulations and data analysis.  However, in the last few years several authors have started questioning this assumption and entertaining the possibility that UHECRs might at least partly consist of nuclei of elements heavier than iron, especially at the highest energies.  Thoroughly testing this hypothesis will require upgrading software so that it can handle such nuclei. In this contribution I will describe the minimal modifications required for the last publicly released version of \SimProp, a code for Monte Carlo simulations of the intergalactic propagation of UHECRs, to be able to treat heavier nuclei, and discuss the applicability of approximations first introduced for lighter nuclei.}

\ConferenceLogo{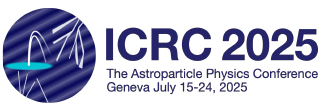}

\FullConference{39th International Cosmic Ray Conference (ICRC2025)\\
 15–24 July 2025\\
Geneva, Switzerland\\}

\begin{document}
\maketitle

\section{Introduction}
Until recently, it was commonly assumed that ultra-high-energy cosmic rays (UHECRs) include no or negligible amounts of nuclei heavier than iron-56.  This assumption was often hard-coded into software for UHECR simulations and data analyses, including the \SimProp~v2r4~\cite{v2r4} Monte Carlo code for the simulation of intergalactic propagation.  Recent works have questioned this assumption, hypothesizing that the most energetic cosmic rays contain considerable fractions of elements heavier than iron \cite{farrar24,zhang24,bortolato24,esmaeili24,FarrarICRC2025,ZhangICRC2025}.  \SimProp\ is undergoing a complete overhaul, \SimProp~Sirente~\cite{sirente}, expected to be released in 2026; in the meantime, in this work I show how to use its latest released version, \SimProp~v2r4, to simulate the propagation of arbitrarily heavy nuclei with only minor modifications to the source code, using cross sections computed by TALYS \cite{talys}.

\section{Command-line options and modifications to the source code}
The default photodisintegration model in \SimProp~v2r4 uses a hard-coded list of nuclides (with one isobar for each mass number~$2 \le A \le 4$~and $9 \le A \le 56$, for a total of 51~nuclides, plus free protons), but when using one of the command-line options \texttt{-M~1}~to \texttt{-M~4} the list of nuclides is read from the standard input (see the \SimProp~v2r4 documentation \cite{v2r4} for more information). 

Furthermore, the $Q$~values of beta decays are read from a hard-coded table which includes nuclides with up to 31 protons and up to 36 neutrons, and computed from the semi-empirical mass formula for nuclides not in the table.  For large mass numbers, the lightest isobar is not always the one predicted by the semi-empirical mass formula, which can cause infinite loops in the code.  However, beta decays can be disabled in \SimProp\ via the \texttt{-D~0}~command-line option, which causes each nuclide to be treated as its respective beta-decay stable isobar.

In addition, even when reading the list of nuclides from the standard input, there are three lines in the C++ source code where the assumption~$A \le 56$ can result in undefined behaviour (such as segmentation faults) or incorrect results for heavier nuclei, and must therefore be edited and the code recompiled before it can reliably be used for heavier nuclei, namely:
\begin{description}
    \item[\texttt{src/SimProp.cc}, l.~64:]
    \texttt{int NatEarth[56] = {0};}
    defining an array which will contain a histogram of the mass numbers of nuclei reaching Earth, resulting in undefined behaviour if any nucleus with~$A > 56$ reaches Earth in the simulation.
    \item[\texttt{src/Output.h}, l.~74:]
    \texttt{static const unsigned nnuc\_max = 100;}
    defining the size of the array which will contain the nuclei reaching Earth for each injected nucleus, resulting in undefined behaviour if more than~100 nuclei reach Earth from the same injected nucleus (e.g.\ in the case of complete photodisintegration of a nucleus with~$A > 100$ into $A$~free nucleons) in the simulation.
    \item[\texttt{src/NucModel.cc}, l.~662:]\!
    \texttt{if(k\_com >= 0 \&\& fA[k\_com]>1 \&\&
fA[k\_com]<56) beta *= (fA[k\_com]/4.);} using the criterion~$1 < A < 56$ to identify nuclei other than protons, and approximating~$Z^2/A$ as~$A/4$ (i.e.~$Z$ as~$A/2$) to scale their electron--positron pair-production energy loss rate compared to that of protons.
\end{description}
For this work, I increased the constants in the first two lines to~512; as for the third line, since $A/2$~is not a particularly good approximation to~$Z$ for nuclei with~$A \gg 40$ to begin with, I replaced the line with \texttt{if (k\_com >= 0) beta *= fZ[k\_com]*fZ[k\_com]/fA[k\_com];} using the actual~$Z$ for any nucleus.

\section{The cross sections}
\begin{wrapfigure}{o}{0.5\columnwidth}
    \begin{verbatim}
        projectile g
        mass 6
        element Li
        energy g0-150.grid
    \end{verbatim}
    \caption{An example input file used with TALYS~2.0}
    \label{fig:input}
\end{wrapfigure}
The 2020 Atomic Mass Evaluation \cite{AME2020} includes measured or estimated atomic masses of nuclides with mass numbers~$1 \le A \le 295$.  For each value of~$A$ between~6 and~295, I selected the isobar with the smallest atomic mass, and for each such nuclide, I used TALYS~2.0 \cite{talys} with default settings (e.g.\ as in \autoref{fig:input}) to compute photodisintegration cross sections for photon energies~$0 < \epsilon' \le 150~\mathrm{MeV}$ in the nucleus rest frame.

For use in \SimProp, I combined the cross sections for the individual photodisintegration channels into a nucleon and an alpha-particle ejection cross section according to the scheme in \cite[eqs.~(4.1) and~(4.2)]{v2r4}, and then fitted them as a Gaussian function for~$\epsilon' < 30~\mathrm{MeV}$ plus a plateau for $30~\mathrm{MeV} \le \epsilon' < 150~\mathrm{MeV}$, as required by the \SimProp\ \texttt{-M~4} command-line option.
Contrary to the claim in the TALYS documentation that it can handle mass numbers up to~$A=339$, for mass numbers $278 \le A \le 285$ it returned NaN values; for these nuclides I used an artificial cross section of $40{,}000~\mathrm{mb}$ to ensure effectively instantaneous disintegration.
The resulting input file for \SimProp
\begin{wrapfigure}[17]{o}{0.5\textwidth}
    \centering
    \includegraphics[width=\linewidth,page=1]{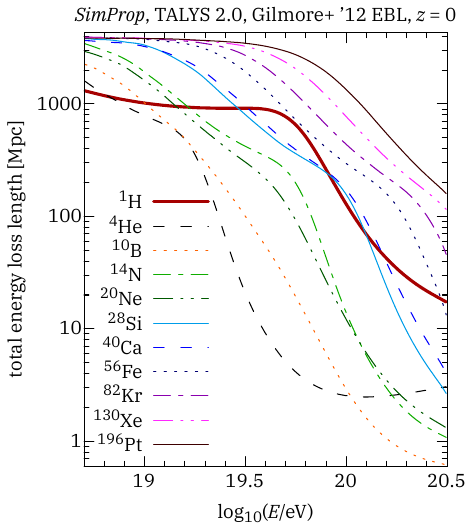}
    \caption{Energy loss lengths for various nuclei}
    \label{fig:rates}
\end{wrapfigure}
\texttt{-M~4} is available at \url{https://baltig.infn.it/adimatteo/superheavy}.

Note that while for certain nuclides the data available for tuning TALYS are rather sparse, certain statistical approaches used in TALYS can be presumed to be more accurate for $A \ggg 1$ than for lighter nuclei.

\section{Example scenarios}
I simulated the injection of pure platinum-196, pure xenon-130, and pure krypton-82 (approximately the peaks of $r$-process nucleosynthesis), with power-law injection energy spectra with the spectral indices indicated in the figures, and a constant source emissivity per unit comoving volume (i.e.\ $\propto (1+z)^{3}$~per unit physical volume).
These scenarios are not necessarily assumed to be astrophysically realistic, and no hypotheses are made about the possible nature or mechanism of sources resulting in them; they are merely chosen as examples of how UHECR nuclei heavier than iron, by being able to travel in intergalactic space for longer distances than even protons, let alone nuclei between hydrogen and iron (\autoref{fig:rates}), can violate some of the common findings about ``conventional'' UHECR nuclei, such as those reported in ref.~\cite{BisterICRC2025}.

For comparison, I also simulated the injection of pure iron-56, pure calcium-40, and pure silicon-28. In all cases, I used the Gilmore \etal\ 2012 model \cite{gilmore12} for the extragalactic background light (EBL) (\texttt{-L~7} command-line option in \SimProp~v2r4).

\subsection{Mass composition at Earth}
\begin{figure}
    \centering
    \includegraphics[width=0.5\textwidth,page=1]{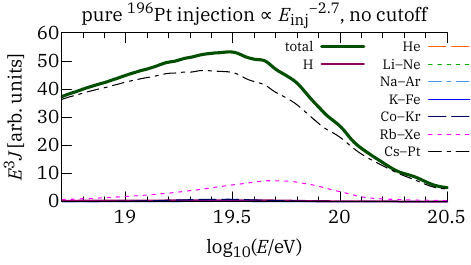}%
    \includegraphics[width=0.5\textwidth,page=4]{spectra.pdf}
    \includegraphics[width=0.5\textwidth,page=2]{spectra.pdf}%
    \includegraphics[width=0.5\textwidth,page=5]{spectra.pdf}
    \includegraphics[width=0.5\textwidth,page=3]{spectra.pdf}%
    \includegraphics[width=0.5\textwidth,page=6]{spectra.pdf}
    \caption{Predicted spectra at Earth from the injection of the three elements heavier than iron (left) and the three ``conventional'' elements (right) with no injection cutoff, grouped by atomic numbers at Earth}
    \label{fig:mass}
\end{figure}
As shown in \autoref{fig:mass}~left, in the cases of injection of nuclei heavier than iron, the vast majority of the injected nuclei reach Earth almost intact, losing at most a handful of nucleons and remaining in the same atomic number bin, and the vast majority of the rest end up in the immediately preceding bin, still retaining more than around half the original nucleons.  Even in the absence of an injection rigidity cutoff, the flux of secondary protons is negligible for platinum and xenon injection and strongly subdominant for krypton injection.  
This contrasts with the cases of injected nuclei lighter than iron, where without an injection cutoff the fluxes at the highest energies would be dominated by secondary protons (\autoref{fig:mass}~right), and so a mass composition at Earth with few or no protons at such energies would indicate the presence of an injection cutoff.

\subsection{Source distances}
\begin{figure}
    \centering
    \includegraphics[width=0.5\textwidth,page=19]{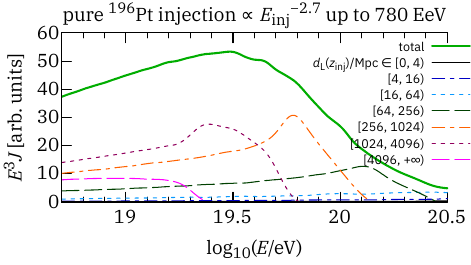}%
    \includegraphics[width=0.5\textwidth,page=22]{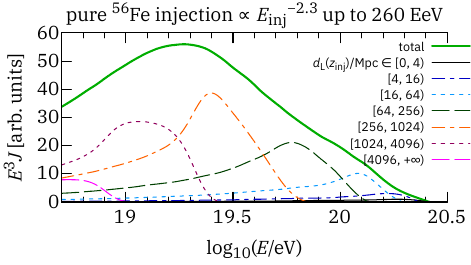}
    \includegraphics[width=0.5\textwidth,page=20]{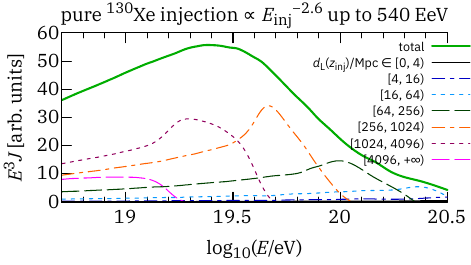}%
    \includegraphics[width=0.5\textwidth,page=23]{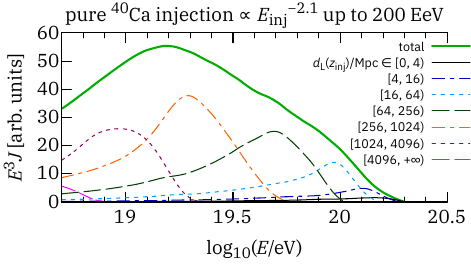}
    \includegraphics[width=0.5\textwidth,page=21]{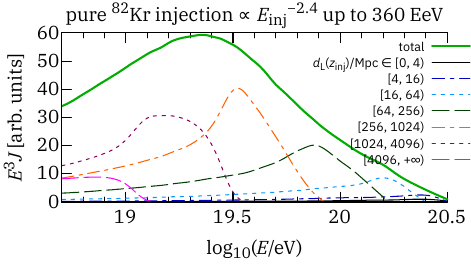}%
    \includegraphics[width=0.5\textwidth,page=24]{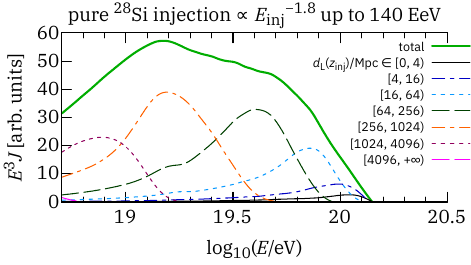}
    \caption{Predicted spectra at Earth from the injection of the three elements heavier than iron (left) and the three ``conventional'' elements (right) with the sharp injection cutoffs indicated in the panel titles, grouped by luminosity distance to the sources}
    \label{fig:dist}
\end{figure}
Catalogues of source candidates used for searches for correlations with UHECR arrival directions typically include objects at distances up to~$250~\Mpc$; the Universe is to a very good approximation isotropic at larger scales.
Furthermore, if only elements up to iron are injected (\autoref{fig:dist}~right), then at the highest energies the propagation lengths are reduced to a few tens of megaparsecs, drastically restricting the number of possible source candidates.  Conversely, in the cases of injection of nuclei heavier than iron (\autoref{fig:dist}~left), the contribution from the local Universe to the flux at Earth is not dominant except at energies above around~60--$100~\EeV$, where the available statistics is too low for correlation studies with large catalogues, and even at the energies of the highest observed events there is a sizeable contribution from sources many tens of megaparsecs away, hampering searches for the sources of individual events.
This extends the ``disappointing model'' trend \cite{disappointing}, whereby the heavier UHECRs are, the harder it is to infer their sources.

\section{Limitations}

Using only one isobar for each mass number disregarding beta decays only produces accurate results if either beta-decay lifetimes are much shorter than all other relevant timescales or the cross sections are approximately independent of the proton number~$Z$ for a fixed mass number~$A$.  The former assumption is known to be false in some cases: certain kinematically possible double beta decays are so slow that they have never been observed, and furthermore neutral atoms can decay by electron capture but fully ionized nuclei in vacuum cannot (though in the CMB they can as a result of Bethe--Heitler pair production \cite{esmaeili24}).

Alpha decays are not implemented in \SimProp, except for~${^8\mathrm{Be}} \to 2\,{^4\mathrm{He}}$ and ${^5\mathrm{He}} \to {^4\mathrm{He}} + n$; therefore, the results may not be very accurate if the propagation of nuclei heavier than lead is simulated.  In addition, the nucleons in a nucleus are treated as independent during pion production, i.e.\ the total cross section is approximated as~$\propto A$.  This approximation is not very accurate for~$A \gg 1$, where a regime intermediate between~$\propto A$ (independent nucleons) and~$\propto A^{2/3}$ (nucleons fully screening each other) can be expected.

Finally, intergalactic magnetic fields (IGMFs) were neglected in this work, but for such large electric charges and propagation distances even modest IGMFs can have a non-negligible impact, suppressing the flux of the lowest-rigidity nuclei.

\section{Conclusions and future directions}
I showed how \SimProp~v2r4 \cite{v2r4} can be used to simulate nuclei heavier than iron with just a few minor modifications.  In this work I only showed results with elements up to platinum, but elements up to lead would presumably result in qualitatively very similar results, as for iron and heavier nuclei the dependence of the energy loss lengths on the nuclear mass is rather regular (\autoref{fig:rates}).  Even nuclei heavier than lead could in principle be simulated, but the results may be less accurate due to the neglect of alpha decay, as well as other approximations.
A new version, \SimProp~Sirente~\cite{sirente}, is expected to be released sometime in 2026, addressing some of these limitations.  The recently released CRPropa~3.3 \cite{CRPropa3.3} also implements nuclei up to lead ($Z=82$).

I showed how findings about light and medium nuclei are not always applicable to heavier nuclei: the absence of protons at Earth at the highest energies would not necessarily imply that there must be an injection cutoff, and the sources may be further away than would be possible for protons, let alone other light or medium nuclei (at least provided that the IGMFs are negligible).

If nuclei from helium to iron are injected, the source emissivity per unit comoving volume cannot have an evolution~$\propto (1+z)^m$ with~$m \gtrsim 3$, because this would result in more secondary nucleons around the ``ankle'' energy~$\approx 5~\EeV$ than observed \cite[and refs.\ therein]{BisterICRC2025}.  Due to their reduced interaction rates, it is not obvious that this also applies to nuclei heavier than iron.  This will be the subject of future studies.

\end{document}